\documentclass[aps,twocolumn,showpacs,floatfix, bibnotes]{revtex4}

\usepackage{graphicx}%
\usepackage{dcolumn}
\usepackage{color}
\usepackage{amsmath}
\usepackage{here}

\begin{document}
\begin{sloppy}

\newcommand{\be}{\begin{equation}}
\newcommand{\ee}{\end{equation}}
\newcommand{\bea}{\begin{eqnarray}}
\newcommand{\eea}{\end{eqnarray}}
\newcommand\bibtranslation[1]{English translation: {#1}}
\newcommand\bibfollowup[1]{{#1}}

\newcommand\pictc[5]{\begin{figure}
                       \centerline{
                       \includegraphics[width=#1\columnwidth]{#3.eps}}
               \protect\caption{\protect\label{fig:#4} #5}
                    \end{figure}            }
\newcommand\pict[4][1.]{\pictc{#1}{!tb}{#2}{#3}{#4}}
\newcommand\rpict[1]{\ref{fig:#1}}

\newcommand\leqt[1]{\protect\label{eq:#1}}
\newcommand\reqtn[1]{\ref{eq:#1}}
\newcommand\reqt[1]{(\reqtn{#1})}

\newcounter{Fig}
\newcommand\pictFig[1]{\pagebreak \centerline{
                   \includegraphics[width=\columnwidth]{#1}}
                   \vspace*{2cm}
                   \centerline{Fig. \protect\addtocounter{Fig}{1}\theFig.}}

\title{Effect of microscopic disorder on magnetic properties of metamaterials}

\author{Maxim Gorkunov$^{1,2}$, Sergey A. Gredeskul$^{1,3}$, Ilya V. Shadrivov$^1$, and Yuri S. Kivshar$^1$}

\affiliation{$^1$Nonlinear Physics Centre, Research School of
Physical Sciences and Engineering, Australian National University,
Canberra ACT 0200, Australia\\
$^2$Institute of Crystallography, Russian Academy of Science, Moscow 119333, Russia\\
$^3$Department of Physics, Ben-Gurion University, Beer-Sheva,
Israel}

\begin{abstract}
We analyze the effect of microscopic disorder on the macroscopic
properties of composite metamaterials and study how weak
statistically independent fluctuations of the parameters of
structure elements can modify their collective magnetic response
and left-handed properties. We demonstrate that even a weak
microscopic disorder may lead to a substantial modification of the
metamaterial magnetic properties, and 10\% deviation in the
parameters of the microscopic resonant elements may lead to a
substantial suppression of wave propagation in a wide frequency
range. A noticeable suppression occurs also if more than 10\% of
resonant magnetic elements possess strongly different properties,
but in this latter case the defects can create an additional weak
resonant line. These results are of the key importance for characterizing and optimizing the novel composite metamaterials with left-handed properties at terahertz and optical frequencies.
\end{abstract}

\pacs{42.70.Qs, 41.20.Jb, 78.67.Pt}

\maketitle

\section{Introduction}

Fabricated composite conductive structures for electromagnetic waves
or {\em metamaterials} acquire a growing attention of researches
during the last years due to their unique properties of magnetic
permeability and left-handed wave propagation. Being composed of
three-dimensional arrays of identical conductive elements, the
metamaterials have much in common with conventional optical crystals
scaled to support the propagation of microwave or Terahertz
radiation. In contrast to crystals, metamaterials allow tailoring
their macroscopic properties by adjusting the type and geometry of
the structural elements. In particular, it is possible to obtain
negative permeability values in magnetically resonant metamaterials
at the frequencies up to hundreds of Terahertz~\cite{NRexp, tera}.
This appears to be especially useful for the practical realization
of the negative refraction phenomenon~\cite{Pendry}.

The simplest method to create a magnetically resonant metamaterial
is to assemble a periodic mesh of the resonant conductive elements
(RCEs), where each element is much smaller than the radiation
wavelength, and it can be well approximated by a linear LC contour.
Introduction of a small slit provides the contour with a certain
capacitance while its shape determines the self-inductance. As a
result, a resonance of induced currents and corresponding magnetization
resonance occurs. Remarkably, the resistive losses in RCEs are low
enough to provide the quality factor of the resonance of the order
of $10^3$~\cite{NRexp}.

During recent years, a number of ideas have been suggested for
achieving better characteristics~\cite{Tretyakov}, tunability
\cite{PRB, BG, Reynet}, nonlinear wave coupling~\cite{PRlow,
ZSK3,PRE} in composite metamaterials by inserting different types of
active and passive electronic complements into the resonant
circuits. One of the common features of these seemingly different
approaches is to modify the macroscopic properties by identical
insertions into the microscopic structure of the metamaterial
assuming a technologically ideal fabrication process. It was always
assumed that all RCEs are identical, and no analysis of the effect
of random deviations or disorder was carried out. However, we should
expect that even weak parameter fluctuations in the microscopic
parameters may become critical near the magnetic resonance. On the
other hand, after a certain operation time, a small amount of RCEs
could experience a breakdown operating in a way different from other
elements. Thus, the problem of disorder appears naturally in the
theory of composite metamaterials.

Recently, the study of defect elements on transmission properties
was studied in experiments with one- and two-dimensional
metamaterial structures~\cite{Zhao}. The very first study of the
effect of disorder in the microscopic structure of metamaterials has
been made in Ref.~\cite{Disorder}, where the magnetic susceptibility
for a spatially uniform system~\cite{EPJ} has been averaged with
respect to random variations of the RCE resonant frequency, and the
resulting change of the frequency dispersion of the left-handed
composite system has been found analytically. The method used in
Ref.~\cite{Disorder} is based on the macroscopic averaging performed
prior to the statistical averaging. This is possible only under the
assumption that the resonant frequency is a slowly varying random
function of coordinate and its correlation length $r_c$ satisfies the inequality
\begin{equation}
    \label{applicability1}
r_\text{resp} \ll r_c \ll  \lambda,
\end{equation}
where $r_\text{resp}$ is the characteristic length of the local
response (which is usually of the order of a few lattice
constants~\cite{EPJ}), and $\lambda$
is the wavelength of the electromagnetic radiation propagating in
the metamaterial. However, the opposite case
\begin{equation}
    \label{applicability2}
    r_c\ll  r_\text{resp},
\end{equation}
appears to be more realistic because even the nearest neighboring
RCEs are statistically independent. Then the primary characteristic
of the problem is not the average susceptibility itself but the
current distribution, and the magnetic properties of the disordered
system are determined by a macroscopic average of the current. If
the averaging length is large with respect to the correlation
length, i.e., if the inequality (\ref{applicability2}) is fulfilled,
then the macroscopic current is a self-averaged quantity~\cite{LGP},
and it should coincide with its ensemble average. In this case, the
magnetic properties of the system are described by a statistical
mean current.

Below we study systematically the effect of disorder on the
averaged characteristics and susceptibility of the metamaterials,
and consider {\em two practically important models} of the
disordered composite metamaterials assuming the capacitances
$C_{{\bf n}}$ of different RCEs to be random quantities. In the
first model, we assume that the capacitances are completely
uncorrelated, i.e. the inequality (\ref{applicability2}) is
automatically satisfied, but the fluctuations are weak. The second
model corresponds to small volume density $\tilde n$
of inserted defect RCEs acting as impurities with substantially different
capacitance $\tilde C$. The difference can be very strong here
covering two practically important cases of casual RCE breakdown,
$\tilde C\rightarrow \infty$, and absence of some RCEs, $\tilde
C\rightarrow 0$. The impurities make the medium microscopically
inhomogeneous. The volume in which the local response is formed
has to contain a large number of impurities. In accordance with
the condition (\ref{applicability2}), this implies an additional
but not very restrictive condition: concentration volume density
of impurities should not be extremely small, $\tilde n \gg
\lambda^{-3}$.

The paper is organized as follows. In Sec.~\ref{sec2} we develop
an appropriate theoretical method based on the response function
of the discrete composite medium
, which allows describing the properties and response of rather
general microscopically inhomogeneous media. Section~\ref{sec3} is
devoted to the study of effects of small deviations in each
resonant element. In Sec.~\ref{sec4}, we deal with the case of
strong but rarefied impurities. Our general conclusions are
accompanied by some particular examples calculated numerically for
the typical metamaterial parameters. Finally, in Sec.~\ref{sec5}
we discuss the quality and reliability requirements for electronic
components to be used as RCE insertions in the composite magnetic
metamaterials.

\section{Response function of magnetic metamaterials}
\label{sec2}

First we consider an ideal composite metamaterial created by a
three-dimensional lattice of identical RCEs. The RCEs are placed in
the parallel planes normal to the $z$ axis (see Fig.~1), and they
form a three-dimensional structure. We denote $N$ the total
macroscopic number of the lattice sites with the index ${\bf n}$.
For the external fields oscillating with the frequency $\omega$, the
relation between the external electromotive forces, $\mathcal{E}
_{\bf n}$, applied to RCEs, and the induced currents $I_{\bf n}$ is
given by {\em the mutual impedance matrix}~\cite{Landau} $\hat Z $
with $N^2$ elements:
\begin{equation}
\label{multiimp}
    \sum_{{\bf m}}Z_{\bf n m} \ I_{\bf m}=\mathcal{E} _{\bf n}.
\end{equation}
where $Z_{{\bf n} {\bf m}}=Z({\bf n}-{\bf m})$ for a periodic
structure.
The diagonal elements of the matrix $\hat Z$ coincide with the RCE
self-impedance:
\begin{equation} \label{selfimp}
  Z(0) = -i\omega L+\frac{i}{\omega C}+R,
\end{equation}
while the non-diagonal ones are determined by the mutual inductance:
\begin{equation}  \label{mutualimp}
  Z({\bf n}-{\bf m}) = -i\omega M({\bf n}-{\bf m})
\end{equation}

\pict[0.7]{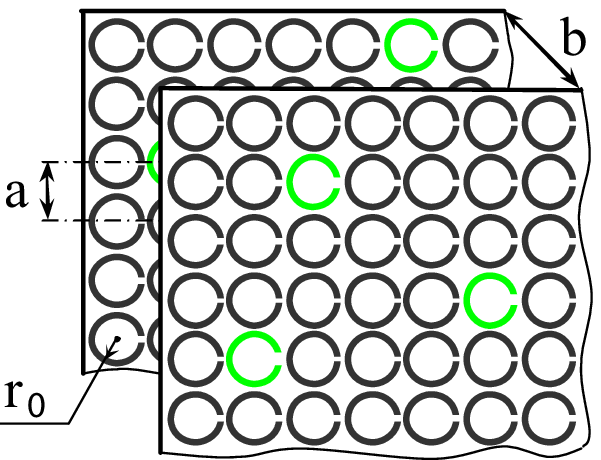}{gork_fig0}{Schematic of two layers of the
metamaterial with random insertions of defect resonant elements (shown in green).}

We define the discrete Green's function of the composite
metamaterial as the distribution of currents $G(\bf n)$ when only one RCE with the index ${\bf n=0}$ is exposed to the
action of the unitary external electro-motive force, i.e. $\mathcal{E}
_{\bf n}=\delta_{\bf n,0} $. Then, Eq.~(\ref{multiimp}) yields
\begin{equation} \label{Green's def}
    \sum_{{\bf m}}Z({\bf n - m}) G({\bf m - l})=\delta_{\bf n,l}
\end{equation}
for arbitrary indices $\bf n$ and $\bf l$.
By the other words, discrete Green's function is the inverse
matrix of $\hat Z$: $\hat Z\hat G=\hat I.$

The inversion of the difference matrix can be easily performed via
the Fourier transform in the reciprocal space. The latter consists
of $N$ wave vectors $\bf{k}$ within the first Brillouin zone. We
define  the Fourier transform  $\tilde{f}({\bf k})$ of a discrete
function $f({\bf n})$ as
\begin{equation*}
    \tilde{f}({\bf k})=\frac{1}{N}\sum_{{\bf n}}f({\bf n})e^{-i{\bf k}{\bf n}}
\end{equation*}
so that the inverse transform has the form
\begin{equation*}
    f({\bf n})=\sum_{{\bf k}}\tilde{f}({\bf k})e^{i{\bf k}{\bf n}}.
\end{equation*}

For the Fourier spectra, using convolution theorem, Eq.~(\ref{Green's def}) reduces to a
simple form $Z({\bf k}) G({\bf k}) =1$, that allows us to write {\em
the discrete Green's function} as follows,
\begin{equation}
\label{Green}
    G({\bf n})=\frac{1}{N}\sum_{{\bf k}}\frac{e^{i{\bf k}{\bf n}}}
    {\displaystyle{\sum_{{\bf m}}Z({\bf m})e^{-i{\bf k}{\bf m}}}}.
\end{equation}

The poles of the Green's function [i.e. zeros of the denominator
in Eq.~(\ref{Green})] determine the spectrum of linear waves that
can be excited in the composite metamaterial.
In the magnetostatic approximation, only a part of the
spectrum can be revealed. Conventional relativistic light-like
electromagnetic waves remain beyond the validity of this approach.
However, as we demonstrate below, small wave vectors and high group
velocities of the light-like waves make their contribution
negligible. The excitations which determine the Green's function on
a microscopic scale can be well explored in the magnetostatic
approximation. Being first mentioned in Ref.~\cite{EPJ} as
Biot-Savart excitons, these linear magnetostatic excitations were
named soon {\em magnetoinductive waves}~\cite{Katya1, Katya2}.
However, these waves can be treated, at least formally, as a
classical analogue of linear excitations of spin systems or magnons,
and we use the term magnons below in this paper. The spectra of
magnons were thoroughly studied in one- and two-dimensional
structures. The spectrum of a three-dimensional structure was
explored only in the nearest and next-nearest neighbor
approximations~\cite{Katya1, Domains}, which provide only a
qualitative information. As it will be discussed in Sec.~\ref{sec3b} below, in order to obtain a quantitatively reliable results we should take into
account hundreds of neighbors.

After changing the summation over the macroscopic number of $\bf
k$-vectors by the integration over the first Brillouin zone, $B_1$,
we obtain
\begin{equation}
G({\bf n})= \frac{1}{(2\pi)^3 n}\  \int\limits_{B_1} \frac{e^{i \bf{k n}} \  d^3
k}{\displaystyle{\sum_{{\bf m}}Z (\bf m)e^{-i\bf k m}}},
\end{equation}
where $n$ is the volume density of RCEs. With the help of
Eqs.~(\ref{selfimp}), (\ref{mutualimp}) we can rewrite the
denominator in the following way
\begin{equation} \label{Gint}
G({\bf n})= \frac{i \omega C}{(2\pi)^3 n}\
\int\limits_{B_1} \frac{\Omega^2
({\bf k})\ e^{i \bf{k n}}\ d^3
k}{\omega^2 - \Omega^2 ({\bf k})+i\omega\gamma({\bf k})}.
\end{equation}
Here the dispersion $\Omega({\bf k})$ is defined as
\begin{equation} \label{Omega}
\Omega^2 ({\bf k})=  \frac{\omega_0^2}{1+ L^{-1}\sum\limits_{\bf
n}M({\bf n}) e^{i \bf k n}},
\end{equation}
and the resonance width is
\begin{equation}
\gamma({\bf k})=\frac{\Omega^2({\bf k})}{Q \omega_0},
\end{equation}
where $\omega_0=(LC)^{-1/2}$ and $Q=\omega_0 L/R$ are the RCE
resonant frequency and quality factor, respectively.

To isolate the contribution of singular poles, we split the integral
in Eq.~(\ref{Gint}) into the real and imaginary parts,
\begin{equation}
\label{GAB}
G({\bf n})= \frac{\omega C}{(2\pi)^3 n}\ \left[ A({\bf n}) + i B({\bf
n})\right],
\end{equation}
where the imaginary part allows a straightforward numerical
integrating,
\begin{equation}\label{B}
B({\bf n}) = \int\limits_{B_1} \frac{\Omega^2({\bf k}) \left[
\omega^2 - \Omega^2 ({\bf k})\right] \ e^{i {\bf k n}}\ }{\left[
\omega^2 - \Omega^2 ({\bf k})\right] ^2+\omega^2 \gamma^2({\bf k})}\
d^3 k,
\end{equation}
whereas the real part
\begin{equation}\label{A}
A({\bf n}) = \int\limits_{B_1} \frac{\omega \ \Omega^2({\bf k}) \gamma({\bf k})
\ e^{i {\bf k n}}\ }{\left[ \omega^2 - \Omega^2 ({\bf k})\right] ^2+\omega^2
\gamma^2({\bf k})}\ d^3 k
\end{equation}
has the contribution at the surface $S_{\bf K}$ built by the magnon wave vectors ${\bf K}(\omega)$ obeying the relation $\Omega({\bf
K}(\omega))=\omega$. Since the RCE quality factor is high, the
singular part dominates in Eq.~(\ref{A}). In the vicinity of the
surface $S_{\bf K}$, we write $\Omega({\bf k})\simeq \omega + {\bf k
\cdot v(K)}$, where ${\bf v(K)}$ stands for the group velocity.
Next, we integrate along the surface normal reducing the volume
integration in Eq.~(\ref{A}) to the surface integral over $S_{\bf
K}$,
\begin{equation} \label{Asurf}
A({\bf n})=\pi \omega \int\limits_{S_{\bf K}} \frac{ e^{i \bf{K n}}\ d^2
K}{v({\bf K})},
\end{equation}
which is now more suitable for numerical calculations.

Analyzing Eqs.~(\ref{B}), (\ref{Asurf}), it is easy to conclude that
accounting for the light-like modes would not lead to any noticeable
corrections. These modes are located in the center of the Brillouin
zone near the point ${\bf k} =0$ and their contribution to the integral $B$ is
apparently small. On the other hand, the light group velocity is about two orders of magnitude higher than that of the magnons. Therefore, the light-like
part in $A$ is also negligible.

\section{Weak fluctuations}
\label{sec3}

\subsection{Modified magnetic permeability}
\label{sec3a}

In order to model the effect of weak disorder in the composite
media, we assume that the values of the RCE self-impedances
experience random uncorrelated deviations due to the capacitance
fluctuations. This may correspond to the results of a real
fabrication process when the capacitances of different resonators
are not identical, and they can be treated as independent
quantities. In this case the correlation radius 
coincides with the lattice constant, $r_c=a$, and the inequality
(\ref{applicability2}) is satisfied. Obviously, strong uncertainty
should totally destroy the macroscopic metamaterial response. We
assume here that the fluctuations are weak and study their effect
on the permeability in the second order of the perturbation theory
with respect to capacitance fluctuations, using the methods known from the solid state physics~\cite{LKTs}.

We define the local fluctuations of the RCE impedance as
\begin{equation}
    \delta_{{\bf n}}=\frac{1}{\omega}\left(\frac{1}{C_{{\bf
    n}}} - \frac{1}{C} \right),
\end{equation}
where
\begin{equation}
    \frac{1}{C} \equiv \bigg< \frac{1}{C_{\bf n}}\bigg>,
\end{equation}
where angle brackets stand for statistical averaging.
To calculate the magnetic permeability, we follow the procedure
described in Ref.~\cite{EPJ} and study the metamaterial exposed to a
homogeneous external field ${\bf H}^{(0)}$. Then
Eq.~(\ref{multiimp}) takes the form,
\begin{equation}\label{electromotive}
    \sum_{{\bf m}}Z({\bf n}-{\bf m})I_{{\bf m}}+i\delta_{{\bf n}}I_{\bf
n}=\mathcal{E},
\end{equation}
where $\mathcal{E}=i \mu_0 \omega S H_z^{(0)}$, and $S$ is the RCE
area. Iterating Eq.~(\ref{electromotive}), we obtain
\begin{multline}\label{In}
    I_{{\bf n}}=I_0-i\sum_{{\bf m}}G({\bf n}-{\bf m})\delta_{{\bf
    m}}I_0-\\
    \sum_{{\bf m},{\bf p}}G({\bf n}-{\bf m})G({\bf m}-{\bf p})\delta_{{\bf
    m}}\delta_{{\bf p}}I_{{\bf p}},
\end{multline}
where $I_0$ is the current induced in RCEs without parameter fluctuations,
\begin{equation}\label{I0}
    I_0= \mathcal{E}\sum_{{\bf n}}G({\bf
    n})=\frac{\mathcal{E}}{Z_0}, \ \ \ Z_0\equiv
    \sum_{\bf n}Z({\bf
    n}).
\end{equation}
The assumption of weak fluctuations allows us to substitute $I_0$
instead of $I_{{\bf p}}$ into the last term of Eq. (\ref{In}).

The important step in our subsequent analysis is macroscopic
averaging of this equation. The size of the (macroscopic) volume of
averaging should be small with respect to the wavelength of the
external field. On the other hand, this size is much larger than the
radius of the capacitance correlations. Therefore, from the
statistical point of view, the volume of averaging can be considered infinite. As a result, instead of volume averaging the
statistical averaging can be performed~\cite{LGP}. Taking into
account that $\langle \delta_{{\bf n}}\rangle=0$, and the
capacitances of different RCEs are statistically independent,
$\langle\delta_{{\bf m}}\delta_{{\bf p}}\rangle \propto \delta_{{\bf
m, p}}$ we obtain that in the second order of the perturbation
theory the average current induced in RCEs can be presented as [cf.
Eq.~(\ref{I0})]
\begin{equation} \label{Ieff}
    \langle I_{{\bf n}} \rangle= \frac{\mathcal{E}}{Z_\text {eff}},
\end{equation}
where the effective impedance
\begin{equation}
    Z_\text{eff}=Z_0+\delta^2 G({\bf 0}),
\end{equation}
involves the square of the standard deviation $\delta^2=\langle
\delta_{\bf n}^2 \rangle$.

After obtaining the effective impedance, we can perform the rest of
the procedure described in Ref.~\cite{EPJ} and finally write the
magnetic permeability as follows,
\begin{equation}\label{mueff}
\mu_{zz}=\dfrac{i Z_\text{eff}- (2/3) \omega \mu_0 n S^2}{i
Z_\text{eff}+ (1/3) \omega \mu_0 n S^2}.
\end{equation}

We note that the fluctuations contribute to both imaginary and
real parts of $Z_\text{eff}$. The imaginary part of the correction
is determined by $B({\bf 0})$ and it affects mainly the real part of the
permeability. The real part of the correction leads to an increase
of $\text{Im}(\mu_{zz})$, i.e., enhances the effective dissipation.
Remarkably, this occurs even when RCE losses are absent, and the
energy is dissipated without conversion into heating, i.e. this
dissipation mechanism is analogous of the Landau damping. The nature
of this non-heating dissipation becomes clear if we note that it
is determined by the term $A({\bf 0})$, which combines the
contributions from all magnetostatic waves excited at the given frequency
$\omega$. This suggests that the incident radiation experiences
scattering on the microscopic fluctuations. Similar effect arising
in electromagnetic media without positional order (see, e.g.
Ref.~\cite{TretBook} and references therein) is known as scattering
loss or Raleigh scattering. In our case, very small correlation
radius suppresses the scattering into light-like modes, but
the scattering into magnons is strong.

\subsection{Weak disorder in a typical metamaterial}
\label{sec3b}

\pict{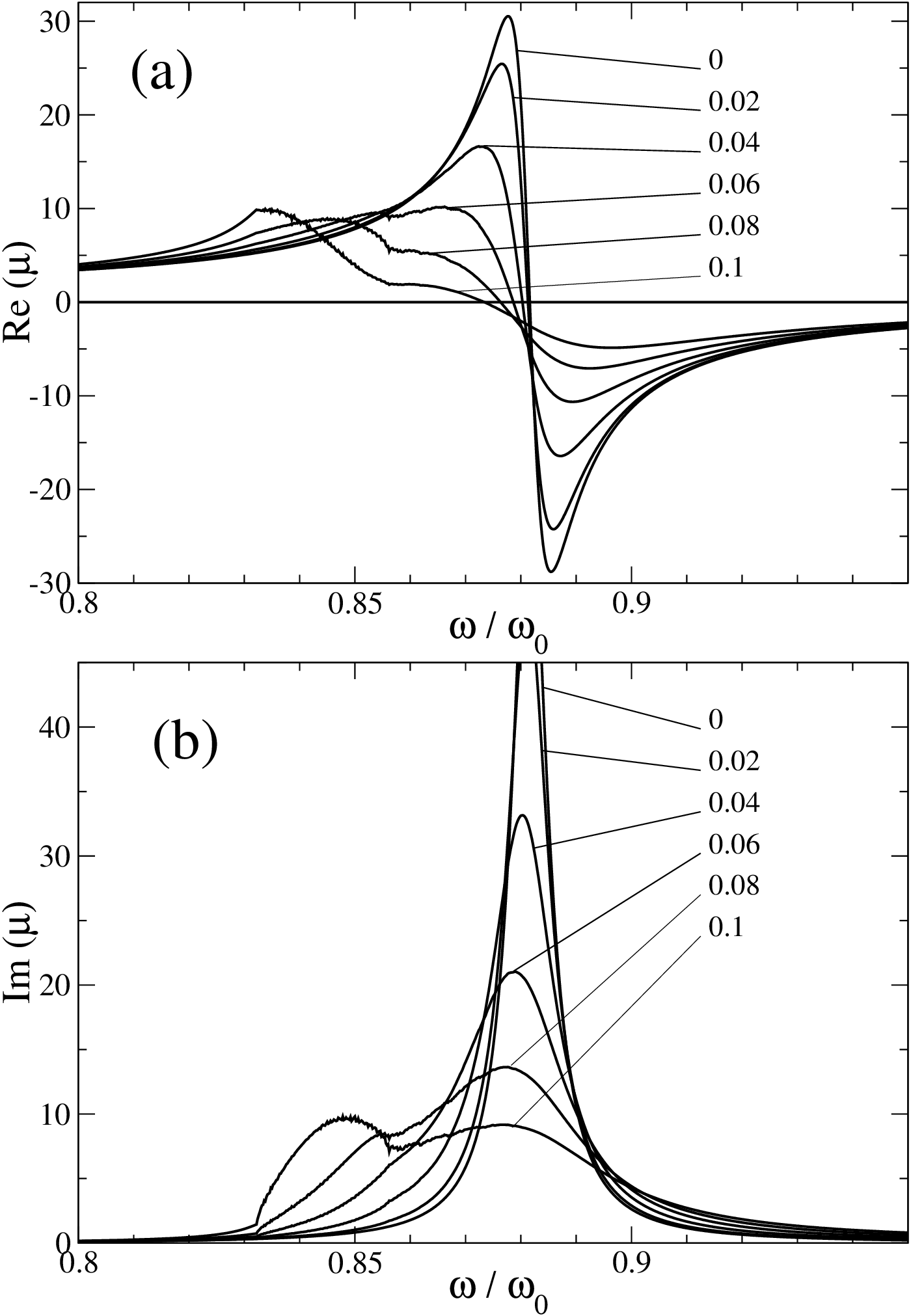}{gork_fig1}{Real (a) and imaginary (b) parts of the
magnetic permeability of metamaterial without fluctuations and with
weak capacitance fluctuations, $\delta \omega_0 C$, marked on the
plots. }

As a particular example of our theory, we calculate the averaged
magnetic permittivity numerically for typical metamaterial
parameters with a weak disorder. We assume circular RCEs with radius
$r_0=2 \text{mm}$, wire thickness $l=0.1~\text{mm}$, which leads to
the self-inductance $L=8.36~\text{nHn}$ (see Ref.~\cite{Landau}). To
obtain RCEs with the resonant frequency $ \omega_0=6 \pi \cdot 10^9
\text{rad/s}$ ($\nu_0=3 \text{GHz}$), we take $C=0.34~\text{pF}$.
The lattice constants are $a=2.1 r_0$ in the plane and $b=r_0$ in
the $z$ direction. The RCE quality factor $Q$ can reach the values
of $10^3$~\cite{NRexp}. However, we expect that insertion of diodes
or other electronic components can lower this value to $Q=10^2$.

First, we calculate the linear spectrum of magnon waves from
Eq.~(\ref{Omega}) and find a strong evidence of long-range
interaction effects: the lattice sum converges rather slow. In
particular, in order to get an accuracy of a few percent, we have to
expand the summation radius to at least ten lattice constants.
Further increase of the summation limits would be unjustified since the
sum should be calculated over the distances much smaller than the
wavelength of radiation. We believe that the problem of exact calculation of
magnon spectrum in the three-dimensional case needs a separate
detailed consideration. On the other hand, we observe that the
maximum uncertainty in the spectrum takes place along specific
directions perpendicular to the edges of the Brillouin zone.
Therefore, the resulting error in the integrals determining the
discrete Green's function is extremely small.

Evaluating numerically the integrals $A(\bf 0)$ and $B(\bf 0)$
according to Eqs.(\ref{B}), (\ref{Asurf}) we obtain the effective
impedance and corresponding magnetic permeability. The frequency
dispersion of the permeability $\mu_{zz}$ is presented in
Fig.~\rpict{gork_fig1} for the unperturbed metamaterial and for several
values of the standard deviation. Apparently, already a 10\%
uncertainty changes dramatically the permeability frequency
dispersion near the resonance. The effect is most pronounced at the
frequencies below the resonance. In a wide range, the imaginary part
of $\mu$ becomes comparable with the real part, and the weakly
disordered nondissipative medium becomes strongly dissipative. In
the range of negative $\mu$ above the resonance, the losses are also
considerably higher than in a perfect metamaterial. As a result, the
frequency range appropriate for the negative refraction shrinks.

\section{Rarefied strong defects}
\label{sec4}

\subsection{Concentration expansion}
\label{sec4a}

We consider now the metamaterial with a small amount of randomly
distributed defective RCEs which differ strongly from the regular
RCEs, and therefore can be treated as impurities. We assume that
the dimensionless concentration of impurities $c=\tilde{n}/ n$ is low, i.e. $c\ll 1$. As above, we assume that the
deviations from the structure parameters appear due to the
difference in RCE capacitances, and the capacitance of the
impurity, ${\tilde C}$, can even become very large corresponding
to a casual breakdown. We neglect any correlation between the
impurities but assume that two impurities cannot be placed
on the same site. The applicability condition, $\tilde n\gg
\lambda^{-3}$, defines the lower limit for the impurity
concentration, $c_{min} \sim 10^{-3}$ for a typical metamaterial.
Above this limit, we can calculate statistically averaged current
and construct its concentration expansion using standard techniques~\cite{Lifshits,LGP}.

In the system with impurities, the microscopic current distribution
is substantially inhomogeneous. We are interested in the normalized
averaged value defined as
\begin{equation}
    \langle I \rangle=\frac{1}{N}\sum_{{\bf n}}I_{\bf n},
\end{equation}
where the summation is performed over a volume containing a
macroscopic number of impurities. The concentration expansion of the
averaged current can be written in the following
form~~\cite{Lifshits,LGP}
\begin{multline}\label{concexp}
    \langle I \rangle=I_0+c \sum_{{\bf p}_1}\left[I_1({\bf
p}_1)-I_0 \right]
    +\\
    \frac{c^2}{2}
    \sum_{{\bf p}_1\neq{\bf p}_2}\left[I_2({\bf p}_1,{\bf p}_2)-
    I_1({\bf p}_1)-I_1({\bf p}_2)+I_0 \right] + \ldots .
\end{multline}
Here
\begin{equation}
\label{one-imp}
    I_1({\bf p}_1)\equiv \frac{1}{N}\sum_{{\bf n}}I_{\bf n}^{(1)}({\bf
p}_1)
\end{equation}
is the averaged current if a single impurity is located at the
site ${\bf p}_1$, while
$$
I_2({\bf p}_1,{\bf p}_2)\equiv \frac{1}{N}\sum_{{\bf n}} I_{\bf
n}^{(2)}({\bf p}_1,{\bf p}_2)
$$
is that for two impurities placed at the sites ${\bf p}_1,{\bf
p}_2$. The terms of higher order involve the averaged solutions
for more impurities. Clearly, we can write formally the
expressions for any term of the expansion \cite{Lifshits}. However,
in this paper we focus on the strongest effects which are linear
in the impurity concentration. To calculate the corresponding
coefficient, we should find first the corresponding current
distribution.

The impedance matrix equation (\ref{multiimp}) in an impure
system exposed to a homogeneous external field takes the form
\begin{equation}
    \sum_{{\bf m}}Z({\bf n}-{\bf m})I_{{\bf m}}+
    i\Delta \sum_{{\bf p}}\delta_{{\bf n, p}} \ I_{{\bf p}}= \mathcal{E},
    \label{DynEq}
\end{equation}
where the summation in the second term in the l.h.s. is performed
over the impurity sites ${\bf p}$ only, and
\begin{equation}
    \Delta=\frac{1}{\omega}\left( \frac{1}{\tilde C}-\frac{1}{C}
    \right).
\end{equation}
Inverting the matrix $\hat Z$, we rewrite Eq.~(\ref{DynEq}) as
follows,
\begin{equation}\label{Iimp}
    I_{{\bf n}}=I_0-i\Delta\sum_{{\bf p}}G({\bf n}-{\bf p})I_{{\bf p}}.
\end{equation}

If a single impurity is located at the site ${\bf p}_1$,
Eq.~(\ref{Iimp}) taken at ${\bf n}={\bf p}_1$ yields
\begin{equation}
    I_{{\bf p}_1}=\frac{I_0}{1+i\Delta G({\bf 0})}
\end{equation}
and the current distribution in this case is
\begin{equation} \label{I1}
    I_{{\bf n}}^{(1)}({\bf p}_1)=I_0\left[1-
    \frac{i\Delta G({\bf n}-{\bf p}_1)}
    {1+i\Delta G({\bf 0})}\right].
\end{equation}
Averaging this expression according to Eq.(\ref{one-imp}) and
substituting the result into Eq.~(\ref{concexp}) leads to the same
results (\ref{Ieff}) and (\ref{mueff}) for the averaged current
and the magnetic permeability correspondingly, where the effective impedance is
\begin{equation}\label{Zeff2}
Z_\text {eff}=Z_0 + c \ \frac{i\Delta}{1+i \Delta G({\bf 0})}.
\end{equation}

\subsection{Effective magnetic permeability}
\label{sec4b}

To illustrate our results with a specific example, we take typical
parameters of metamaterials used above in Sec.~\ref{sec3b}.
First, we study the effect of impurities with an infinite
capacitance, $\tilde C \rightarrow \infty$, which corresponds, for
instance, to a casual breakdown of varactor diode insertions. Next,
we deal with other impurities which either do not contribute to
the magnetization or are just absent; this latter case is modelled by
setting $\tilde C \rightarrow 0$. Although these cases are two
opposite extreme limits, the resulting averaged permeabilities look
similar and differ only by few percents. This result is not
surprising, because in both cases the currents induced in the
defect RCEs are either small or vanish.

\pict{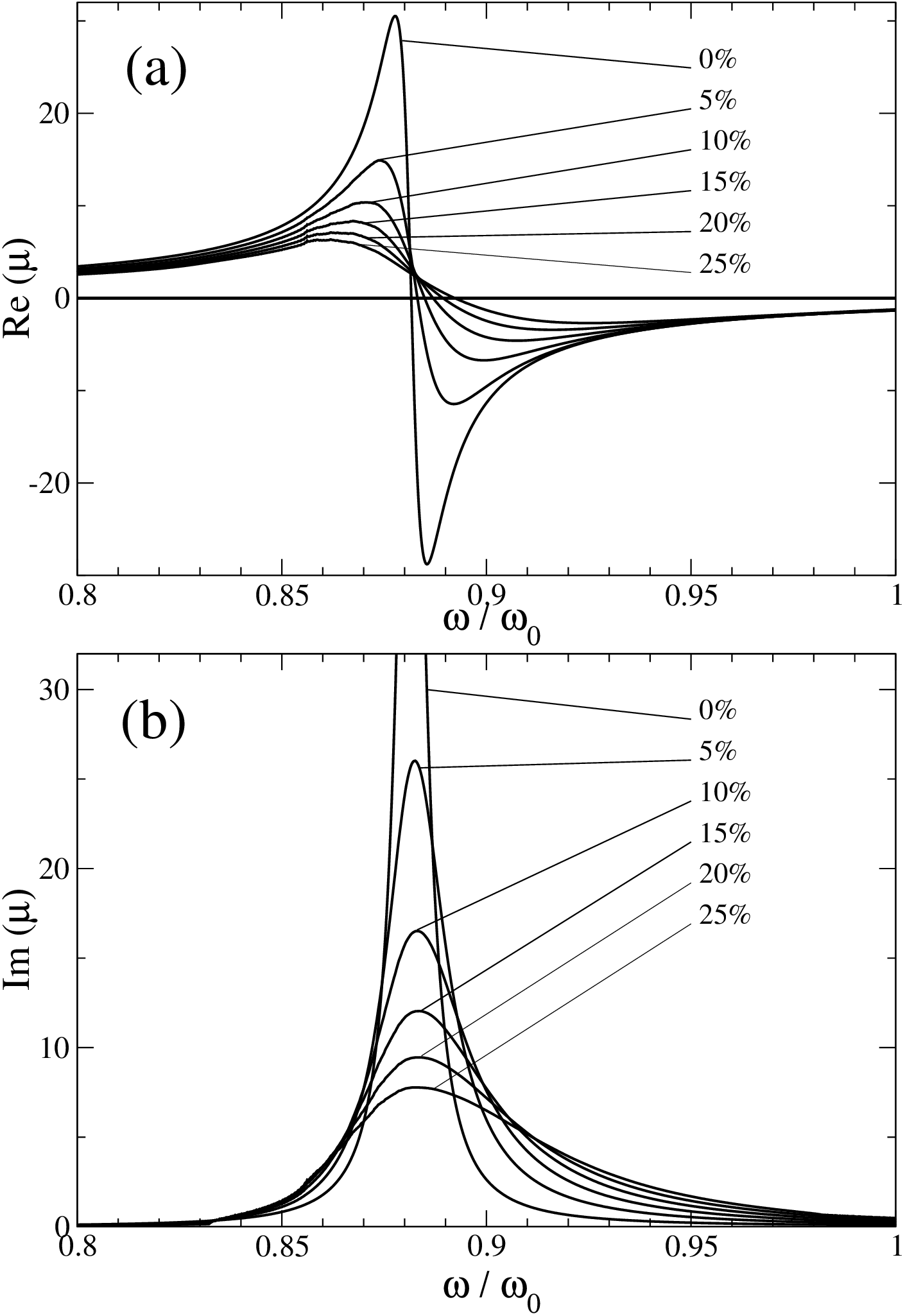}{gork_fig2}{(a) Real and (b) imaginary parts of the
magnetic permeability of metamaterials without and with impurities
(impurity concentration is shown near the curves).}

In Fig.~\rpict{gork_fig2} we present the real and imaginary parts of
the magnetic permeability for several concentrations of missing
RCEs. As follows from those results, the resonance is attenuated
much more than one could expect from the fact that a small part of
RCEs do not contribute to the resonance. The reasons for that
enhancement are scattering losses which decrease the effective
quality factor by an order of magnitude or more. We see also that
the metamaterial is more sensitive to such perturbations in the
frequency range of negative $\mu$. The composite medium can tolerate
5\% or less malfunctioning RCEs, but for 10\% and more a substantial
damping of the resonance occurs, and the imaginary part of the
permeability becomes comparable with the real one.

The impurities with a finite resonant frequency, $\tilde \omega_0
\neq 0$, which strongly differs from the main frequency $\omega_0$,
cause observable effects even if their concentration is extremely
low. As seen in Fig.~\rpict{gork_fig3}, only 1\% of such impurities
can build up their own weak resonance, with the position of the
resonance shifted from the impurity eigenfrequency.

In the linear approximation used here, the effect of mutual interaction
of impurities is neglected. Therefore, we are not accounting for the
magnons arising at frequencies close to the impurity resonance. We
expect that the corresponding scattering losses will broaden the
resonance. However, the detailed analysis of this and related
phenomena is beyond the scope of this paper.

\pict{fig04}{gork_fig3}{(color online) Real (solid) and
imaginary (dashed) parts of the permeability of metamaterial with
1\% of randomly distributed defects with the eigenfrequency
$\tilde{\omega}_0=0.5 \omega_0$. The inset shows a blow up of the impurity resonance.}

\section{Concluding remarks}
\label{sec5}

We have demonstrated that even a weak microscopic disorder in the
conducting elements of composite metamaterials can lead
to a substantial modification of their resonant magnetic response.
According to our results, already 10\% disorder in the parameters
causes substantial scattering of the incident radiation into
magnetization waves. This modifies strongly the macroscopic
permeability of the metamaterial leading to the increase of losses. We
believe the study of disorder and effective losses is of a critical
importance for engineering novel metamaterials with various
electronic elements inserted into their microscopic resonators. The
inserted elements should possess no more than a few percent
deviation of their capacitance and/or inductance to ensure that the
metamaterial properties are not distorted dramatically. Another
restriction concerns the insertion stability. The data obtained
demonstrate that the metamaterial tolerates no more than 10\%
disorder in the parameters of the resonant conductive elements.
Casual breakdown of a greater part of the insertions causes
strong damping of the wave propagation in metamaterials.

Our results that even a small amount of defects can build up a
noticeable additional magnetic resonance look useful for suggesting
a simple methods for the sensitive quality control of metamaterials.
If an RCE contains not a single capacitive element but a combination
of insertions with an ordinal slit, a casual insertion breakdown
would switch the RCE to another resonant frequency. The
corresponding narrow gap (or peak) in the metamaterial transmission
can be used as a sensitive indicator of the concentration of damaged
RCEs.

Finally, we point out that the effects of microscopic disorder would
be crucially important in the nanostructured metamaterials even
with simple RCEs without electronic components. Clearly, the fabrication of resonant elements on such scales is less accurate, so that random
fluctuations of the RCE shapes can lead to deviations of the
self-inductance and capacitance. Disorder in position of RCEs
determines the deviations of mutual inductance. We expect that the
results obtained here can be important for qualitative estimations
of nanostructured left-handed media. However, the problem requires a
separate and more systematic analysis.

\section*{Acknowledgments}

This work was supported by the Australian Research Council. Maxim
Gorkunov and Sergey Gredeskul thank Nonlinear Physics Centre of
the Australian National University for a warm hospitality and
financial support during their stay in Canberra.

\end{sloppy}
\end{document}